\documentclass[aps,twocolumn,prl,showpacs]{revtex4}
\usepackage{graphicx}
\begin{document}
\draft
\preprint{19 October 2004}
\title{Effects of edges in spin-$\frac{1}{2}$ bond-alternating
       Heisenberg chains:\\
       Matrix-product variational approach}
\author{Kei-ichi Funase and Shoji Yamamoto}
\address{Division of Physics, Hokkaido University,
         Sapporo 060-0810, Japan}
\date{19 October 2004}
\begin{abstract}
We make a matrix-product variational approach to spin-$\frac{1}{2}$
ferromagnetic-antiferromagnetic bond-alternating chains with anisotropy
on their ferromagnetic bonds, especially under the open boundary
condition.
The rich phase diagram containing the Haldane, large-$D$, and two types of
N\'eel phases is well reproduced with only two variational parameters.
The on-bond anisotropy has a significant effect on the ferromagnetic
coupling between neighboring spins and induces novel edge states peculiar
to spin-$\frac{1}{2}$ chains.
\end{abstract}
\maketitle

   Haldane's conjecture \cite{H464,H1153} sparked renewed interest in
low-dimensional quantum magnets and led to extensive explorations of spin
gaps$-$energy gaps in magnetic excitation spectra.
While the early research centered on spin-$1$ antiferromagnetic chains,
spin-$\frac{1}{2}$ bond-alternating chains have been attracting further
interest due to their enriched ground-state properties
\cite{H2207,H8268,Y9555,H847,O2587,S251}.
A spin-$\frac{1}{2}$ Heisenberg chain with alternating ferromagnetic and
antiferromagnetic couplings converges to the spin-$1$ antiferromagnetic
Heisenberg chain as the ferromagnetic coupling tends to infinity, where
the ground state remains unique and gapful for all the ratios of the two
exchange couplings \cite{H2207}.
Thus the spin-$1$ Haldane gap turns out to be continuously connected to
the gap originating in decoupled singlet dimers.
Such a consideration can be verified for existent bond-alternating
chain compounds such as
IPACuCl$_3$
($\mbox{IPA}=\mbox{isopropylammonium}=(\mbox{CH}_3)_2\mbox{CHNH}_3$)
\cite{B125} and
(4-BzpipdH)CuCl$_3$
($\mbox{4-BzpipdH}
 =\mbox{4-benzylpiperidinium}=\mbox{C}_{12}\mbox{H}_{18}\mbox{N}$)
\cite{R2603}.
These materials indeed reproduce many of observations common to
Haldane-gap antiferromagnets \cite{M564,H1792,M3913,M675}.
With increasing temperature, the effective spin-$1$ features disappear
into the spin-$\frac{1}{2}$ paramagnetic behavior
\cite{M14279,M144428,M2694}.

   Apart from paramagnetic spin $\frac{1}{2}$'s, quantum spin-$\frac{1}{2}$
degrees of freedom lie in spin-$1$ Haldane-gap antiferromagnets
\cite{W2863,M1459,M913,W3844,Y9528,W10345,K420,Y4327,Y3649,S16115,T3956}.
In more general, there appear fractional spin-$\frac{S}{2}$
degrees of freedom on boundaries of antiferromagnetic Heisenberg chains
with integral spin $S$ at low temperatures \cite{Y4051,Y3364}.
The chain-end fractional spins are understandable in view of the
valence-bond-solid states \cite{A799,A477} and were actually observed for
both spin-$1$ \cite{H3181,A2786,A460} and spin-$2$ \cite{Y6831}
antiferromagnetic chain compounds.
Such observations are peculiar to the Haldane phase.
Hence both quantum and classical spin-$\frac{1}{2}$ degrees of freedom
may be observed for Haldane-gap materials composed of spin
$\frac{1}{2}$'s.
Thus motivated, we study spin-$\frac{1}{2}$
ferromagnetic-antiferromagnetic bond-alternating chains with particular
emphasis on their edge states, making good use of the matrix-product
representation as well as a quantum Monte Carlo method.

   The Hamiltonian of our interest is given by
\begin{eqnarray}
   &&
   {\cal H}
   =\sum_{j=1}^{N}
    \bigl[
    -J_{\rm F} \mbox{\boldmath$S$}_{2j-1}\cdot\mbox{\boldmath$S$}_{2j}
    +J_{\rm AF}\mbox{\boldmath$S$}_{2j}\cdot\mbox{\boldmath$S$}_{2j+1}
   \nonumber\\
   &&\qquad\quad
    +D(S_{2j-1}^z+S_{2j}^z)^2
    \bigr],
   \label{E:H}
\end{eqnarray}
where $J_{\rm F}$ and $J_{\rm AF}$ are both set positive.
The on-bond anisotropy $D$, which is related to a single ion anisotropy
in the case of spin $1$, originates from possible dipole-dipole and/or
anisotropic exchange interactions between the ferromagnetically coupled
spin $\frac{1}{2}$'s \cite{M14279}.
When we consider a wave function of the matrix-product type:
\begin{equation}
   |\Psi\rangle
   ={\rm Tr}[g_1\otimes g_2\otimes\cdots\otimes g_N],
   \label{E:Psi}
\end{equation}
employing the complete set of states on the $j$th ferromagnetic bond:
\begin{eqnarray}
   &&
   |s\rangle_j
   =\frac{1}{\sqrt{2}}
    (|\uparrow\downarrow\rangle_j-|\downarrow\uparrow\rangle_j),\ \ 
   |t_-\rangle_j=|\downarrow\downarrow\rangle_j,
   \nonumber\\
   &&
   |t_0\rangle_j
   =\frac{1}{\sqrt{2}}
    (|\uparrow\downarrow\rangle_j+|\downarrow\uparrow\rangle_j),\ \ 
   |t_+\rangle_j=|\uparrow\uparrow\rangle_j,
\end{eqnarray}
the decoupled dimers for $J_{\rm F}\rightarrow 0$ are
described by
\begin{equation}
   g_j
   =\left(
    \begin{array}{cc}
     |s\rangle_j+|t_0\rangle_j & -\sqrt{2}|t_+\rangle_j
     \\
      \sqrt{2}|t_-\rangle_j & |s\rangle_j-|t_0\rangle_j
    \end{array}
    \right),
   \label{E:gDD}
\end{equation}
while the optimum ground state for $J_{\rm F}\rightarrow\infty$ is given
by
\begin{equation}
   g_j
   =\left(
    \begin{array}{cc}
     |t_0\rangle_j & -\sqrt{2}|t_+\rangle_j
     \\
     \sqrt{2}|t_-\rangle_j & -|t_0\rangle_j
    \end{array}
    \right).
   \label{E:gVBS}
\end{equation}
The $g$ matrices (\ref{E:gVBS}), with their indices contracted, create
singlet bonds in between \cite{T1639} and end up as the spin-$1$
valence-bond-solid state \cite{A477}.
The $g$ matrices (\ref{E:gDD}) can be used as the basis for
a variational calculation \cite{B7161} of the Hamiltonian (\ref{E:H}) by
allowing different amplitudes for the singlet and triplet contributions
and reducing the rotational invariance as
\begin{equation}
   g_j
   =\left(
    \begin{array}{cc}
     b|s\rangle_j+c|t_0\rangle_j &
     -\sqrt{2}a|t_+\rangle_j
     \\
      \sqrt{2}a|t_-\rangle_j &
     b|s\rangle_j-c|t_0\rangle_j
    \end{array}
    \right).
   \label{E:gval}
\end{equation}
The isotropic chain is described by equalizing $c$ to $a$ and the two
extreme cases (\ref{E:gDD}) and (\ref{E:gVBS}) are indeed included in this
ansatz.
Then, eq. (\ref{E:Psi}) gives a variational ground state of the periodic
chain, while each of the four elements of the $2\times 2$ matrix
$g_1\otimes g_2\otimes\cdots\otimes g_N$ corresponds to a ground state of
the open chain with fixed spins at each end.
In the Haldane phase, the four states of the open chain are
quasi-degenerate \cite{K5737}.

\begin{figure}
\centering
\includegraphics[width=80mm]{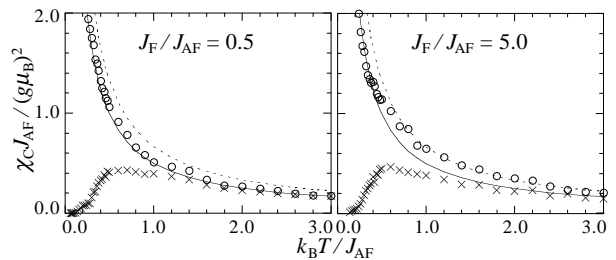}
\vspace*{-3mm}
\caption{Quantum Monte Carlo calculations of the Curie component in the
         magnetic susceptibility as a function of temperature for the
         isotropic ($D=0$) open chains with ferromagnetic ($\circ$) or
         antiferromagnetic ($\times$) couplings at each end.
         The solid and dotted lines denote the Curie susceptibilities due
         to two spins of $S=\frac{1}{2}$ and a single spin of $S=1$,
         respectively.}
\label{F:chiC}
\vspace*{-1mm}
\end{figure}

   The variational energy
$\langle\Psi|{\cal H}|\Psi\rangle/\langle\Psi|\Psi\rangle
 \equiv E_{\rm var}$
is calculated as
\begin{eqnarray}
   &&
   \frac{E_{\rm var}}{2N}
  =-J_{\rm F}
    \frac{2|a|^2-3|b|^2+|c|^2}{4(2|a|^2+|b|^2+|c|^2)}
   \nonumber\\
   &&\qquad
   -J_{\rm AF}
    \frac{2|a|^2(|b|^2+|c|^2+3|b||c|)
         +|a|^4+|b|^2|c|^2}
         {(2|a|^2+|b|^2+|c|^2)^2}
   \nonumber\\
   &&\qquad
   +D
    \frac{2|a|^2}
         {2|a|^2+|b|^2+|c|^2},
\end{eqnarray}
and its bounds are compared with quantum Monte Carlo findings in
Table \ref{T:Eg}.
\begin{table}
\vspace*{-1mm}
\caption{The matrix-product variational (MP) and quantum Monte Carlo (QMC)
         calculations of the ground-state energy per site for the
         isotropic ($D=0$) chain.}
\label{T:Eg}
\begin{ruledtabular}
\begin{tabular}{ccc}
 $J_{\rm F}/J_{\rm AF}$ & MP & QMC \\
\hline
$0.2$ & $-0.37678$ & $-0.3768(1)$ \\
$0.5$ & $-0.38538$ & $-0.3855(1)$ \\
$1.0$ & $-0.41197$ & $-0.4125(1)$ \\
$2.0$ & $-0.49435$ & $-0.4976(1)$ \\
$5.0$ & $-0.82583$ & $-0.8426(1)$ \\
\end{tabular}
\end{ruledtabular}
\vspace*{-1mm}
\end{table}
\begin{figure}[b]
\vspace*{-1mm}
\centering
\includegraphics[width=80mm]{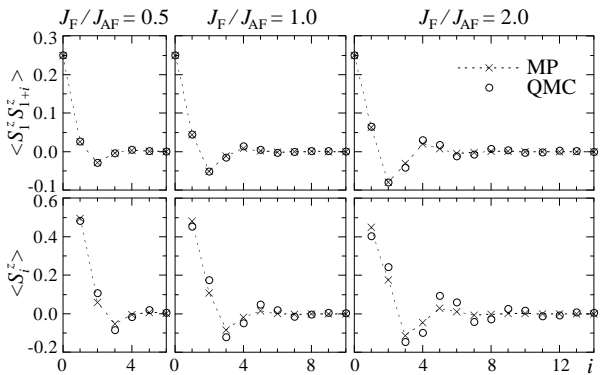}
\vspace*{-3mm}
\caption{The matrix-product variational (MP) and quantum Monte Carlo (QMC)
         calculations of the spin correlations (the upper three) and
         moments (the lower three) in $z$ direction for the isotropic
         ($D=0$) open chains with ferromagnetic couplings at each end.}
\label{F:SS}
\vspace*{-1mm}
\end{figure}
The variational estimates are less precise in the region of strong
ferromagnetic coupling but still agree with the numerical values within
two-percent error even at $J_{\rm F}/J_{\rm AF}=5.0$.
The spin correlation function and local moments induced on a chain end
are also calculated and verified in Fig. \ref{F:SS}.
Considering that the correlation length is considerably underestimated
by wave functions of the matrix-product type in general \cite{K3336},
our variational findings are fairly good.
The variational expression for the $z$-component correlation length $\xi$,
\begin{equation}
   \frac{1}{\xi}
   =\frac{1}{2}
    {\rm ln}\Bigl(\frac{|b|^2+|c|^2+2|a|^2}{|b|^2+|c|^2-2|a|^2}\Bigr),
\end{equation}
is useful especially in the region of weak ferromagnetic coupling.
The edge moments, which are assembled into an effective spin
$\frac{1}{2}$, are characteristic of the Haldane phase \cite{W2863,M913}
and should therefore be observed for the present system with nonmagnetic
impurities as well.
Besides direct observations, we can detect them through magnetic
susceptibility measurements \cite{Y4051,Y3364}.
Figure \ref{F:chiC} shows the Curie component of the susceptibility,
$\chi_{2N+2}^{\rm op.}-\chi_{2N}^{\rm per.}\equiv\chi_{\rm C}$, where
$\chi_L^{\rm per.}$ and $\chi_L^{\rm op.}$ are the susceptibilities of the
periodic and open chains with $L$ spins, respectively.
Two excess moments of $S=\frac{1}{2}$ are found at both low and high
temperatures.
The low-temperature ones are quantum mechanically correlated effective
spins induced at each end, whereas the high-temperature ones are
paramagnetic spins of classical aspect.
In the region of strong ferromagnetic coupling, such a difference can be
understood better, because the distinct features of paramagnetic spin
$1$'s appear at intermediate temperatures, where the Haldane phase is
thermally broken but the ferromagnetic exchange interactions still
survive.
The open chain with antiferromagnetic couplings at each end exhibits no
edge moment at low temperatures even in the Haldane phase.

\begin{figure}
\vspace*{-1mm}
\centering
\includegraphics[width=80mm]{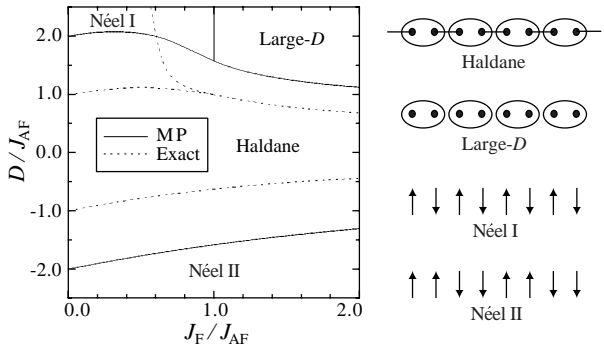}
\vspace*{-3mm}
\caption{The matrix-product variational (MP) and numerical diagonalization
         (Exact) \cite{H8268} calculations of the ground-state phase
         diagram.
         Schematical representations of each phase are presented for
         reference, where the arrow (the bullet symbol) and the segment
         denote a spin $\frac{1}{2}$ with its fixed (unfixed) projection
         value and a singlet pair, respectively, while the circle means
         an operation of constructing a spin $1$ by symmetrizing the two
         spin $\frac{1}{2}$'s inside.}
\label{F:PhD}
\vspace*{-1mm}
\end{figure}

\begin{figure}
\vspace*{-1mm}
\centering
\includegraphics[width=80mm]{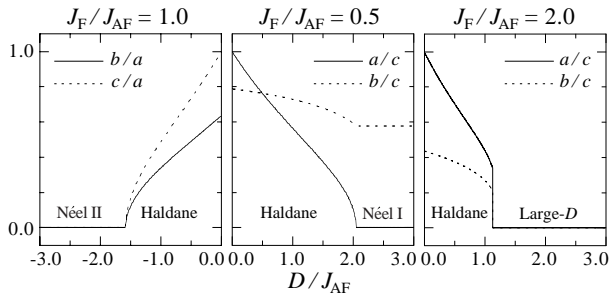}
\vspace*{-3mm}
\caption{The optimum variational parameters as functions of the
         ferromagnetic coupling and the on-bond anisotropy.}
\label{F:VWF}
\vspace*{-1mm}
\end{figure}

   Our variational scheme is still useful for the anistropic chain and
well reproduces its rich phase diagram, as is shown in Fig. \ref{F:PhD}.
With increasing $|D|$, the Haldane gap decreases to zero and there
alternatively appear three distinct phases according to the sign of $D$
and the ratio of $J_{\rm F}$ to $J_{\rm AF}$.
The variational wave function illuminates in itself how the spin
correlations vary with the anisotropy.
Moving across the phase transitions, we plot in Fig. \ref{F:VWF} the
optimum variational parameters.
Considering that eq. (\ref{E:gval}) is rewritten as
\begin{equation}
   g_j=b |s\rangle_j\sigma^0
      +a(|t^x\rangle_j\sigma^x+|t^y\rangle_j\sigma^y)
      +c |t^z\rangle_j\sigma^z,
\end{equation}
where
$(|t^x\rangle_j\pm{\rm i}|t^y\rangle_j)/\sqrt{2}=|t_\pm\rangle_j$,
$|t^z\rangle_j=|t_0\rangle_j$, and
$\sigma^0$ and $(\sigma^x,\sigma^y,\sigma^z)$ are the $2\times 2$ unit
matrix and the Pauli matrices, respectively,
we find that the anisotropy-induced phases are characterized in terms of
$S(S+1)=(\mbox{\boldmath$S$}_{2j-1}+\mbox{\boldmath$S$}_{2j})^2$ and
$S^z=S_{2j-1}^z+S_{2j}^z$ as
i)   Large-$D$: $S=1$,      $S^z=0$;
ii)  N\'eel I :             $S^z=0$;
iii) N\'eel II: $S=1$,      $S^z=\pm 1$.
Thus, the large-$D$ and N\'eel II phases appear in a spin-$1$ chain as
well, while the N\'eel I phase is characteristic of the spin-$\frac{1}{2}$
chain.
No ferromagnetic correlation between any neighboring spins in the N\'eel I
phase can be understood by rewriting the Hamiltonian as
\begin{eqnarray}
   &&
   {\cal H}
   =\frac{1}{2}DN
   +\sum_{j=1}^{N}
    \Bigl[
    -\frac{J_{\rm F}}{2}(S_{2j-1}^+S_{2j}^- + S_{2j-1}^-S_{2j}^+)
   \nonumber\\
   &&\quad
    +(2D-J_{\rm F})S_{2j-1}^zS_{2j}^z
    +J_{\rm AF}\mbox{\boldmath$S$}_{2j}\cdot\mbox{\boldmath$S$}_{2j+1}
    \Bigr].
\end{eqnarray}
With increasing $D$, the ferromagnetic coupling is reduced and turned into
an antiferromagnetic one.

\begin{figure*}
\vspace*{-1mm}
\centering
\includegraphics[width=160mm]{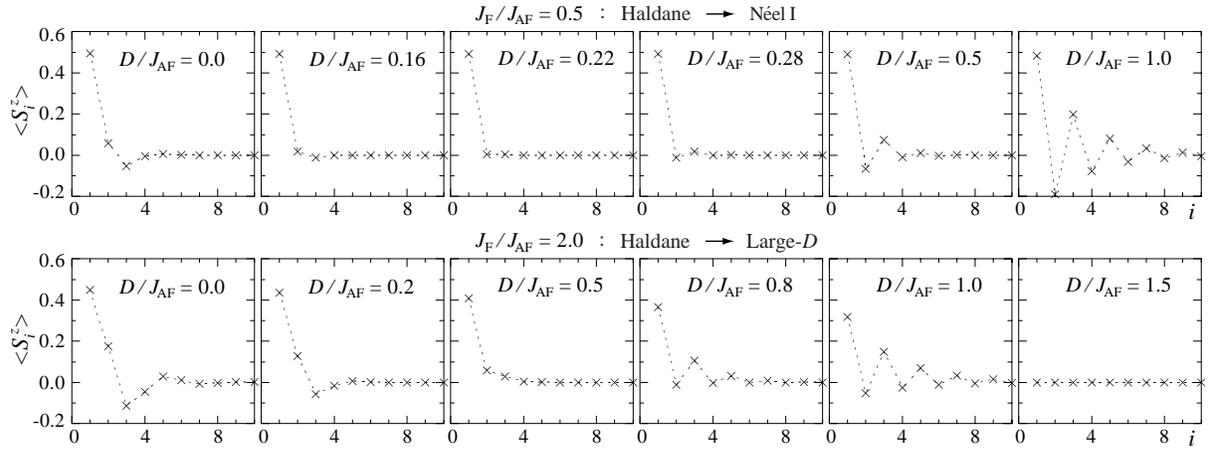}
\vspace*{-3mm}
\caption{The matrix-product variational (MP) calculations of the spin
         moments in $z$ direction for the anisotropic open chains.}
\label{F:Saniso}
\vspace*{-1mm}
\end{figure*}

   We have already observed in Fig. \ref{F:SS} that the spin-$\frac{1}{2}$
open chain also exhibits effective moments of $S=\frac{1}{2}$ at each end.
In the isotropic case, there appears an alignment of the type
$\uparrow\uparrow\downarrow\downarrow\uparrow\uparrow\downarrow\downarrow
 \cdots$
at a chain end.
On the way from the Haldane to N\'eel II phases, such an alignment remains
unchanged and monotonically grows into a long-range order, which is
convincing on the analogy of our experience for a spin-$1$ chain
\cite{Y6277}.
On the ways to the N\'eel I and large-$D$ phases, on the other hand, we
have novel observations peculiar to the spin-$\frac{1}{2}$ chain.
Figure \ref{F:Saniso} shows the edge moments as functions of $D$.
On both the ways from the Haldane to N\'eel I and large-$D$ phases, the
spin alignment changes with increasing $D$ from
$\uparrow\uparrow\downarrow\downarrow\uparrow\uparrow\downarrow\downarrow
 \cdots$ to
$\uparrow\downarrow\uparrow\downarrow\uparrow\downarrow\uparrow\downarrow
 \cdots$ via
$\uparrow\uparrow\uparrow\uparrow\uparrow\uparrow\uparrow\uparrow\cdots$.
The ferromagnetic alignment at a chain end is another interesting effect
of $D$ and is induced near $D/J_{\rm AF}=0.22$ and $D/J_{\rm AF}=0.5$
in the cases of $J_{\rm F}/J_{\rm AF}=0.5$ and $J_{\rm F}/J_{\rm AF}=2.0$,
respectively.
The magnetic moments in $z$ direction form a long-range order in the
N\'eel phases, whereas they vanish in the large-$D$ phase.
The total moment induced at a chain end,
$\sum_{j=1}^{N/2}(S_{2j-1}^z+S_{2j}^z)\equiv S_{\rm edge}^z$, is
variationally expressed as
\begin{equation}
   S_{\rm edge}^z
   =\frac{1}{2}
    \left[
     1-\Bigl(\frac{|b|^2+|c|^2-2|a|^2}{|b|^2+|c|^2+2|a|^2}\Bigr)^{N/2}
    \right].
\end{equation}
The absolute value of the fraction
$(|b|^2+|c|^2-2|a|^2)/(|b|^2+|c|^2+2|a|^2)$ is smaller than unity in the
Haldane phase, while it is fixed to unity in any other phase.
Thus, the effective moment of $S=\frac{1}{2}$ persists in the edge of an
infinite chain throughout the Haldane phase and disappears with the
collapse of the Haldane gap.

   Possible dipole-dipole and/or anisotropic exchange interactions between
ferromagnetically coupled two spins act as a fictitious single-ion
anisotropy \cite{M14279} and high-frequency electron-spin-resonance
experiments on IPACuCl$_3$ \cite{M144428} indeed revealed the
nondegenerate triplet excitation.
Therefore, we take more and more interest in nonmagnetic-ion substitution
at the Cu site of spin-$\frac{1}{2}$ ferromagnetic-antiferromagnetic
bond-alternating chain compounds.
The present system may be compared with further polymerized
bond-alternating copper(II) complexes such as
the trimerized chain compound
3CuCl$_2\cdot$2dx
($\mbox{dx}=\mbox{1,4-dioxane}=\mbox{C}_4\mbox{H}_8\mbox{O}_2$)
\cite{L1403} and the tetramerized chain compound
Cu(3-Clpy)$_2$(N$_3$)$_2$
($\mbox{3-Clpy}=\mbox{3-chloropyridine}=\mbox{C}_5\mbox{ClH}_4\mbox{N}$)
\cite{E4466}, which behave as a spin-$\frac{3}{2}$ critical
antiferromagnet \cite{H2359,O245} and a spin-$(\frac{3}{2},\frac{1}{2})$
ferrimagnet \cite{N214418}, respectively.
With bond alternation, impurity-induced magnetic effects vary with the
location of the dopant impurities in general, as is shown in Fig.
\ref{F:chiC}.
A comparative study on the edge states of various bond-alternating chain
compounds will lead to brand-new observations.

   The authors are grateful to K. Hida, K. Okamoto and T. Sakai for
valuable comments.
This work was supported by the Ministry of Education, Culture, Sports,
Science and Technology of Japan, and the Iketani Science and Technology
Foundation.

\end{document}